\documentclass[conference]{IEEEtran}

\usepackage{amsmath,amssymb,amsfonts}
\usepackage{algorithmic}
\usepackage{graphicx}
\usepackage{svg}
\usepackage{textcomp}
\usepackage{xcolor}
\usepackage{makecell}   %
\usepackage{array}      %
\usepackage{graphicx} 
\usepackage{tabularx}
\usepackage{multirow}

\usepackage{enumitem}
\usepackage{tcolorbox}  %
\usepackage{subcaption}

\usepackage{booktabs}
\usepackage{pifont}  %

\def\BibTeX{{\rm B\kern-.05em{\sc i\kern-.025em b}\kern-.08em
    T\kern-.1667em\lower.7ex\hbox{E}\kern-.125emX}}
\begin{document}

\title{Where Do LLMs Still Struggle? An In-Depth Analysis of Code Generation Benchmarks}

\author{
\IEEEauthorblockN{
Amir Molzam Sharifloo\IEEEauthorrefmark{1}, 
Maedeh Heydari\IEEEauthorrefmark{1}, %
Parsa Kazerooni\IEEEauthorrefmark{1}, 
Daniel Maninger\IEEEauthorrefmark{1}\IEEEauthorrefmark{2}, 
Mira Mezini\IEEEauthorrefmark{1}\IEEEauthorrefmark{2}\IEEEauthorrefmark{3}
}
\IEEEauthorblockA{
\IEEEauthorrefmark{1}Technische Universität Darmstadt, Germany\\
\IEEEauthorrefmark{2}The Hessian Center for Artificial Intelligence (hessian.AI), Germany\\
\IEEEauthorrefmark{3}National Research Center for Applied Cybersecurity ATHENE, Germany\\
}
\IEEEauthorblockA{
amir.molzam@tu-darmstadt.de, \{maedeh.heydari \textbar\ parsa.kazerooni\}@stud.tu-darmstadt.de,\\
daniel.maninger@tu-darmstadt.de, mezini@cs.tu-darmstadt.de
}
}

\maketitle

\begin{abstract}
Large Language Models (LLMs) have achieved remarkable success in code generation, and the race to improve their performance has become a central focus of AI research. Benchmarks and leaderboards are increasingly popular, offering quantitative rankings of LLMs. However, they provide limited insight into the tasks that LLMs consistently fail to solve—information that is crucial for understanding current limitations and guiding the development of more capable models.
To address this gap, we examined code generation tasks across four popular benchmarks, identifying those that major LLMs are most likely to fail. To understand the causes of these failures, we investigated whether the static complexity of solution code contributes to them, followed by a systematic inspection of 114 tasks that LLMs consistently struggled with.
Our analysis revealed four recurring patterns of weaknesses in LLMs, as well as common complications within benchmark tasks that most often lead to failure.

\end{abstract}


\section{Introduction}

Large Language Models (LLMs) are rapidly transforming software development by automating code generation tasks. Trained on massive corpora of open-source code, these systems are now widely adopted in practice---recent surveys indicate that many developers regularly use AI-assisted coding tools \cite{cset}. As the adoption of code-generating LLMs accelerates, rigorous benchmarking becomes increasingly critical. Effective benchmarks provide a standardized framework to evaluate model capabilities, track progress over time, and compare systems on common grounds.

A variety of benchmarks have been proposed to evaluate LLM code generation \cite{humaneval-extended,mbpp-extended,jain2024livecodebench-extended,bigcodebench-extended,swebench-extended}. Early efforts, such as \textit{Most Basic Python Problems (MBPP)} \cite{mbpp-extended} and \textit{HumanEval} \cite{humaneval-extended}, focus on functional correctness for small Python programs, probing models’ understanding of algorithms, language, and basic libraries. Extensions like MBPP+ and HumanEval+ \cite{evalplus} increase test coverage to reveal edge cases, while HumanEval Pro and MBPP Pro \cite{humanevalpro-extended} introduce compositional tasks requiring multiple function calls, testing reasoning across generation steps. Other benchmarks emphasize scale and realism: \textit{APPS} \cite{apps-extended} offers thousands of competitive programming problems with multiple test cases, and \textit{LiveCodeBench} \cite{jain2024livecodebench-extended} continuously updates problems from platforms like LeetCode and AtCoder. \textit{BigCodeBench} \cite{bigcodebench-extended} evaluates use of external libraries across domains, while repository-level benchmarks such as \textit{SWE-bench} \cite{swebench-extended} assess LLM performance on multi-file software development tasks.

Despite the widespread adoption of these benchmarks, there has been little systematic analysis of their tasks and the cases where LLMs failed to generate correct code. Identifying failure-inducing tasks and understanding the underlying causes of the failures is essential for guiding both benchmark design and model development. The prior work \cite{icse2025Wang} analyzed the errors that LLMs made when generating code for the HumanEval benchmark and reported both syntactic and semantic mistakes. However, its scope was limited to HumanEval and did not consider more diverse or challenging benchmarks. In contrast, this paper addresses this gap by conducting an empirical study of popular function-level code generation benchmarks, aiming to answer the following research questions:

\begin{itemize}[left=0pt, label={}, nosep]
\item \textbf{RQ1:} Which tasks do LLMs consistently fail to solve?
\item \textbf{RQ2:} To what extent can these failures be explained by the complexity of the solution code?
\item \textbf{RQ3:} Which additional factors, beyond code complexity, influence LLM failures?
\end{itemize}

To investigate these questions, we conducted a task-level study of four widely used benchmarks (MBPP, HumanEval, BigCodeBench, and LiveCodeBench) and evaluated six representative advanced LLMs selected to reflect diverse and popular code generation scenarios (Claude Sonnet-4, DeepSeek-V3, Qwen3-Coder, GPT-4o, Llama-3.3-70B, and Mistral-3.2-24B). We additionally implemented a tool to measure task complexity from the corresponding solution code, which we used to examine its correlation with model failures. Interestingly, complexity alone did not fully account for these failures, suggesting that additional factors play a significant role. Analyzing the 114 consistently failed tasks across benchmarks, we identified four recurring failure patterns: \textit{wrong problem mapping, flawed or incomplete algorithm design, edge case mishandling, and formatting mistakes}. This analysis reveals how factors beyond solution complexity contribute to LLM failures. We further observed that some benchmark tasks are extremely ambiguous, so that certain failures do not necessarily reflect actual model weaknesses. Our contributions are as follows:

\begin{itemize}[left=0pt, label={}, nosep]
\item \textbf{(1)} Detailed task-level experimental results for four widely used benchmarks evaluated on six major LLMs.
\item \textbf{(2)} A method for measuring solution code complexity and analyzing its correlation with LLM failure rates.
\item \textbf{(3)} Fine-grained failure inspection of 114 tasks, revealing patterns of weaknesses in LLMs and challenges arising from benchmark design.
\end{itemize}
The experimental data and analysis scripts are publicly released on GitHub to support reproducibility and future research.\footnote{\label{note:repo}https://github.com/breath24/FailureBench}.

\section{Experimental Setup}

\textbf{Benchmarks:} We conducted a survey of existing code generation benchmarks and selected those that are both widely adopted in the research community and equipped with executable test cases, enabling reliable evaluation of generated code. Our final selection comprises HumanEval (5,845 citations as of August 2025), MBPP (2,142), LiveCodeBench (524), and BigCodeBench (240), representing a diverse yet widely used set of benchmarks for code generation. In this paper, we focused on benchmarks that specialize in function generation, where a natural language problem is given (with or without contextual code) and the model is required to generate a function that implements the expected functionality.

We used the latest version of each benchmark available to minimize data contamination, thereby increasing the likelihood of observing failures. The APPS benchmark was excluded, as it is widely used for training and would risk contamination. Due to the large number of tasks in MBPP, we relied on a subset of 378 tasks, used in recent extensions such as MBPP+.
For LiveCodeBench, we used LCB-V6, which consists of 175 programming problems released between January and April 2025. For BigCodeBench, we focused on the BCB-Hard subset (148 tasks) using instruct prompts, which contains the most difficult tasks to challenge LLMs. In total, we experimented with 865 tasks across four benchamrks. %

\textbf{Models:} As for the LLMs, we chose six representative advanced LLMs---namely Claude Sonnet-4, DeepSeek-V3, Qwen3-Coder, GPT-4o, Llama-3.3-70B and Mistral-3.2-24B---from different projects that have been widely used. Unfortunately, we excluded Gemini and Grok due to our limited budget and their costly inference fees.

\textbf{Evaluation Procedure:} To identify which tasks lead to failure, we ran the benchmarks on the selected LLMs and collected the generated code. Each task was evaluated using a single generated solution per model (PASS@1): a task was classified as a success if the solution passed all test cases, and as a failure otherwise. For failed tasks, additional details were recorded for further analysis. %
Table~\ref{tab:model_failures} reports the number of failures for each model across the benchmarks.

\begin{table}[h!]
\caption{The number of task failures across benchmarks.}
\centering
\resizebox{\columnwidth}{!}{%
\begin{tabular}{lccc c}
\hline
\textbf{Model} & \textbf{HumanEval} & \textbf{MBPP} & \textbf{LCB} & \textbf{BCB-Hard} \\
\hline
Claude Sonnet-4 & 2  & 11 & 54 & 109 \\
Qwen3-Coder & 7  & 9 & 73 & 102 \\
DeepSeek-V3 & 14 & 9 & 76 & 107 \\
GPT-4o & 25 & 14 & 85 & 107 \\
Mistral-3.2-24B & 28 & 23 & 110 & 114 \\
Llama-3.3-70B & 32 & 32 & 100 & 101 \\
\hline
\end{tabular}}
\label{tab:model_failures}
\end{table}

\textbf{Results:} To provide a categorization of task difficulty based on actual LLM performance, we further classified tasks according to the number of models that failed to solve them, resulting in seven categories ranging from 0 to 6, where 0 represents tasks solved by all models, and 6 represents tasks for which all models failed to generate correct code. Table \ref{tab:all-benchmarks-failures} shows the number of failures in each category.

\begin{table*}[h!]
\caption{Distribution of model failures across benchmarks. For each benchmark, columns 0–6 indicate the number of models that failed a task: 0 = solved by all models, 1 = failed by 1 model, …, 6 = failed by all models. This provides a difficulty categorization of tasks.}
\centering
\resizebox{\textwidth}{!}{%
\begin{tabular}{l|ccccccc|ccccccc|ccccccc|ccccccc}
\hline
\textbf{Model} 
& \multicolumn{7}{c|}{\textbf{HumanEval (164 tasks)}} 
& \multicolumn{7}{c|}{\textbf{MBPP (378)}} 
& \multicolumn{7}{c|}{\textbf{LiveCodeBench (175)}} 
& \multicolumn{7}{c}{\textbf{BCB-Hard (148)}} \\
\hline
\# Models Failed (0–6)
 & 0 & 1 & 2 & 3 & 4 & 5 & 6
 & 0 & 1 & 2 & 3 & 4 & 5 & 6
 & 0 & 1 & 2 & 3 & 4 & 5 & 6
 & 0 & 1 & 2 & 3 & 4 & 5 & 6 \\
\hline
Claude Sonnet-4   
           & 113 & 0 & 0 & 0 & 0 & 1 & 1 
           & 318 & 3 & 0 & 2 & 2 & 2 & 2 
           & 43 & 0 & 2 & 1 & 6 & 10 & 35 
           & 14 & 2 & 4 & 6 & 8 & 13 & 76 \\
Qwen3-Coder
           & 113 & 1 & 0 & 0 & 2 & 3 & 1 
           & 318 & 1 & 1 & 1 & 1 & 3 & 2 
           & 43 & 2 & 3 & 6 & 9 & 18 & 35 
           & 14 & 0 & 2 & 5 & 7 & 12 & 76 \\
DeepSeek-V3
           & 113 & 0 & 3 & 3 & 3 & 4 & 1 
           & 318 & 3 & 0 & 1 & 1 & 2 & 2 
           & 43 & 0 & 7 & 6 & 10 & 18 & 35 
           & 14 & 1 & 2 & 6 & 8 & 14 & 76 \\
GPT-4o    
           & 113 & 4 & 10 & 2 & 4 & 4 & 1 
           & 318 & 4 & 3 & 1 & 1 & 3 & 2 
           & 43 & 1 & 8 & 8 & 15 & 18 & 35 
           & 14 & 2 & 1 & 6 & 7 & 15 & 76 \\
Mistral-3.2-24B 
           & 113 & 6 & 10 & 3 & 4 & 4 & 1 
           & 318 & 13 & 3 & 0 & 2 & 3 & 2 
           & 43 & 7 & 16 & 13 & 18 & 21 & 35 
           & 14 & 3 & 5 & 9 & 7 & 14 & 76 \\
Llama-3.3-70B
           & 113 & 11 & 9 & 4 & 3 & 4 & 1 
           & 318 & 21 & 5 & 1 & 1 & 2 & 2 
           & 43 & 9 & 14 & 8 & 14 & 20 & 35 
           & 14 & 3 & 4 & 4 & 7 & 7 & 76 \\
\hline
\hline
Number of Tasks      
           & 113 & 22 & 16 & 4 & 4 & 4 & 1
           & 318 & 45 & 6 & 2 & 2 & 3 & 2
           & 43 & 19 & 25 & 14 & 18 & 21 & 35
           & 14 & 11 & 9 & 12 & 11 & 15 & 76 \\
\hline
\end{tabular}%
}
\label{tab:all-benchmarks-failures}
\end{table*}

\begin{itemize}[left=0pt]

\item{\textbf{HumanEval:}}
Out of 164 tasks, 113 were solved correctly by every model. There was only one task for which none of the models managed to generate the correct code. Claude Sonnet-4 performed exceptionally good with only two failures, while Llama-3.3-70B records high in failing with tasks that other models were able to solve. %

\item{\textbf{MBPP:}} Throughout the benchmark, 318 tasks were solved by every model. Two tasks were not solved by any of the models. Qwen3-Coder and DeepSeek-V3 emerged as the best models, while Llama-3.3-70B failed the highest number of tasks.

\item{\textbf{LiveCodeBench:}} Among all tasks, 35 could not be solved correctly by any model, while 43 were solved correctly by every model. Claude Sonnet-4 demonstrated the strongest performance with only 54 failures, while Mistral-3.2-24B produced most failures.

\item{\textbf{BCB-Hard:}} In this benchmark, only 14 tasks were correctly solved by all models, while 76 tasks were consistently failed. All models produced many failures with a failure rate range from 68\% to 77\%.
\end{itemize}

\tcbset{
  colframe=black,  %
  colback=white,   %
  coltitle=white,  %
  boxrule=0.8pt,   %
  arc=4pt,         %
  left=2mm,
  right=2mm,
  top=1mm,
  bottom=1mm
}

\begin{tcolorbox}[title={\textbf{RQ1:} Which tasks do LLMs consistently fail at?}]
Our experiments identified 114 tasks across four benchmarks that all models consistently failed. BCB-Hard had the most failures, followed by LiveCodeBench. %
\end{tcolorbox}

\label{sec:method}

\section{Code Complexity Analysis}

Solution code complexity provides a quantitative lens on task difficulty, independent of natural language prompts. To explore whether this complexity correlates to LLM performance, we analyzed ground-truth solutions across benchmarks. We first describe the complexity metrics we use, then report descriptive statistics, followed by correlation and regression analyses to examine their relationship with model failures.

\subsection{Complexity Measurement}

We designed and implemented an algorithm to quantitatively measure the complexity of solution code. We used a set of different code complexity dimensions, each of which captures a distinct facet of what makes a code snippet more cognitively or technically challenging. The dimensions are defined as follows:

\begin{itemize}[left=0pt]
  \item \textbf{Cyclomatic Complexity (CC)}: Measures the intricacy of control flow constructs, including conditionals (\texttt{if}), loops (\texttt{for}, \texttt{while}), and branching. A higher CC value indicates more involved execution paths and decision points.

  \item \textbf{Data Structures}: Assesses the extent and variety of data structure usage, such as lists, dictionaries, sets, and user-defined containers. %

  \item \textbf{Function Calls}: Captures the frequency of function invocation. %

  \item \textbf{Code Length}: Serves as a surface-level proxy for overall complexity, as longer code may correspond to richer logic.

  \item \textbf{Nesting Depth}: Reflects the maximum depth of syntactic nesting, such as loops within conditionals or recursive calls embedded in other structures. %

  \item \textbf{Recursions}: Specifically tracks the presence and frequency of recursive functions. %
\end{itemize}

We use static analysis of abstract syntax trees to compute each of these metrics, averaging values across all tasks in a benchmark. Table \ref{tab:code_metrics} presents the results of these measurements. As the measurement shows, recursions are extremely rare across benchmarks and that the use of data structures is relatively limited. LiveCodeBench demonstrates greater or nearly equal values for all metrics comparing to other benchmarks, highlighting that its solutions require highest code complexity.

\begin{table}[h]
\centering
\caption{Code complexity metrics-average value- across benchmarks.
}
\small
\begin{tabular}{lcccc}
\hline
\textbf{Metric} & \textbf{HumanEval} & \textbf{MBPP} & \textbf{LCB} & \textbf{BCB-Hard} \\
\hline
CC                & 3.34 & 2.58 & 9 & 4.66 \\
Data Structures   & 0.26 & 0.10 & 0.62 & 0.69 \\
Function Calls    & 3.10 & 2.07 & 12.97 & 11.89 \\
Length            & 6.30 & 6.68 & 29.54 & 15.91 \\
Nesting Depth     & 8.37 & 8.70 & 10.48 & 9.42 \\
Recursions        & 0.09 & 0.09 & 0.25 & 0.03 \\
\hline
\end{tabular}
\label{tab:code_metrics}
\end{table}

\subsection{Correlation Analysis}

\begin{figure*}[t]
  \centering
  \begin{subfigure}{0.40\textwidth}
    \includegraphics[width=\linewidth]{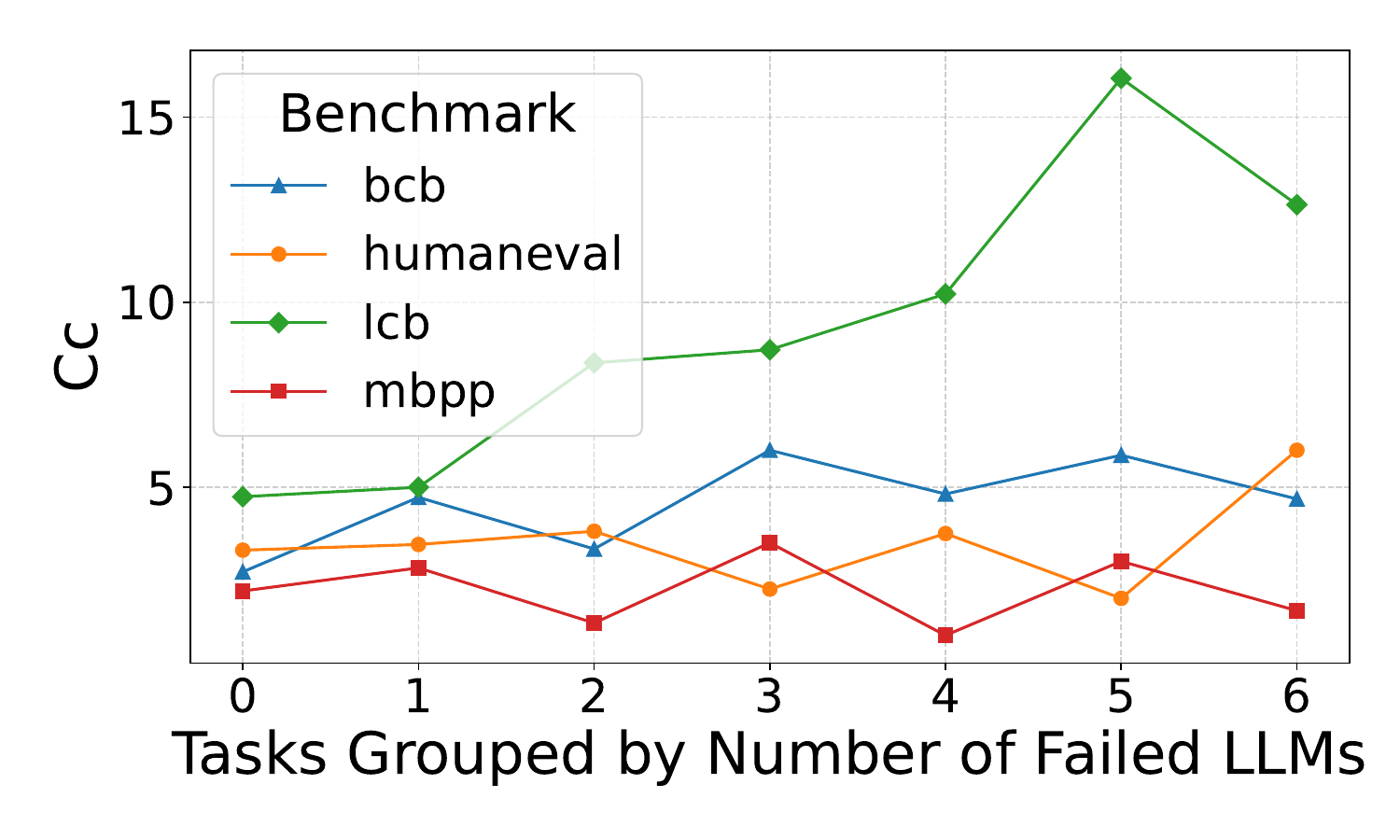}
    \caption{Cyclomatic Complexity}
  \end{subfigure}
  \hspace{1em}
  \begin{subfigure}{0.40\textwidth}
    \includegraphics[width=\linewidth]{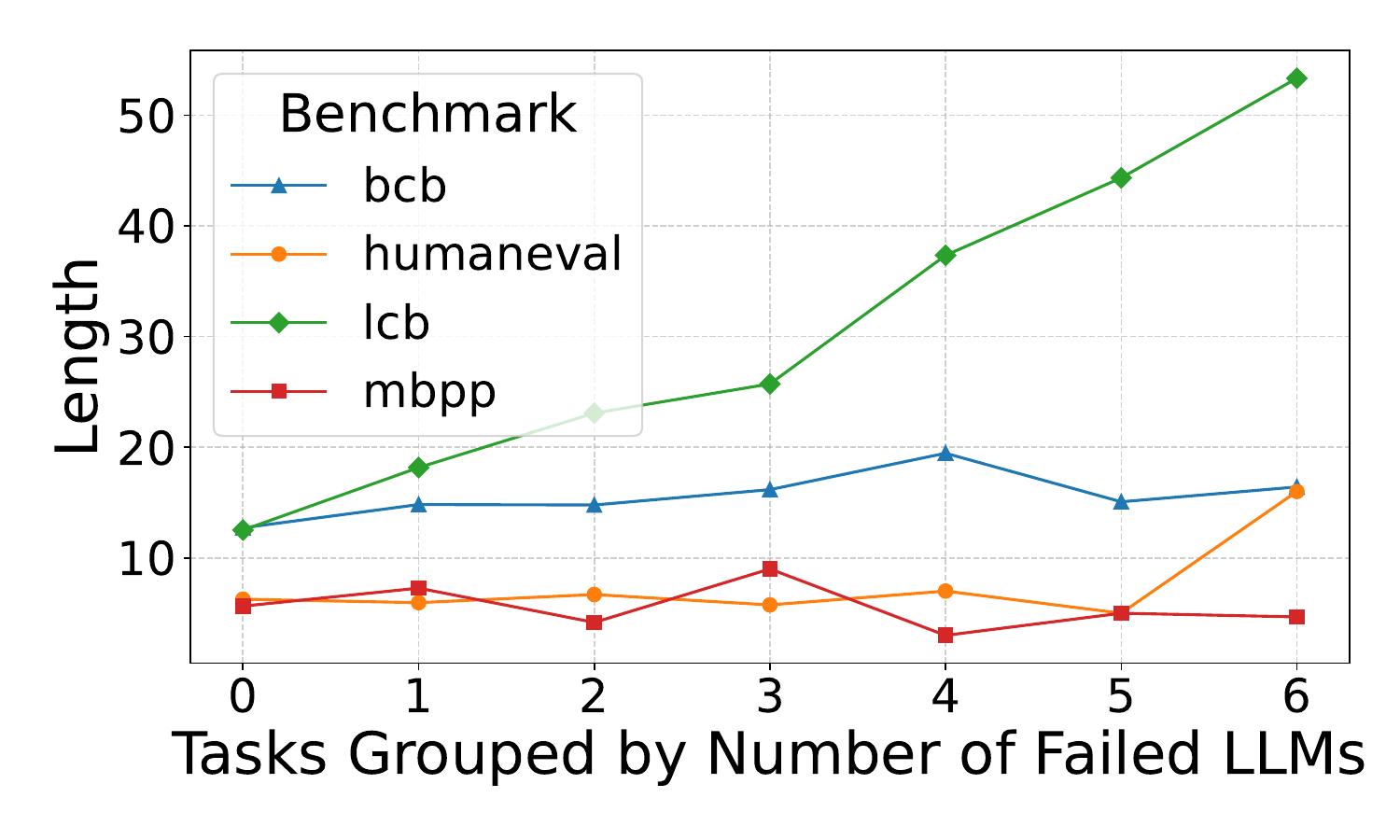}
    \caption{Length}
  \end{subfigure}

  \vspace{1em}

  \begin{subfigure}{0.40\textwidth}
    \includegraphics[width=\linewidth]{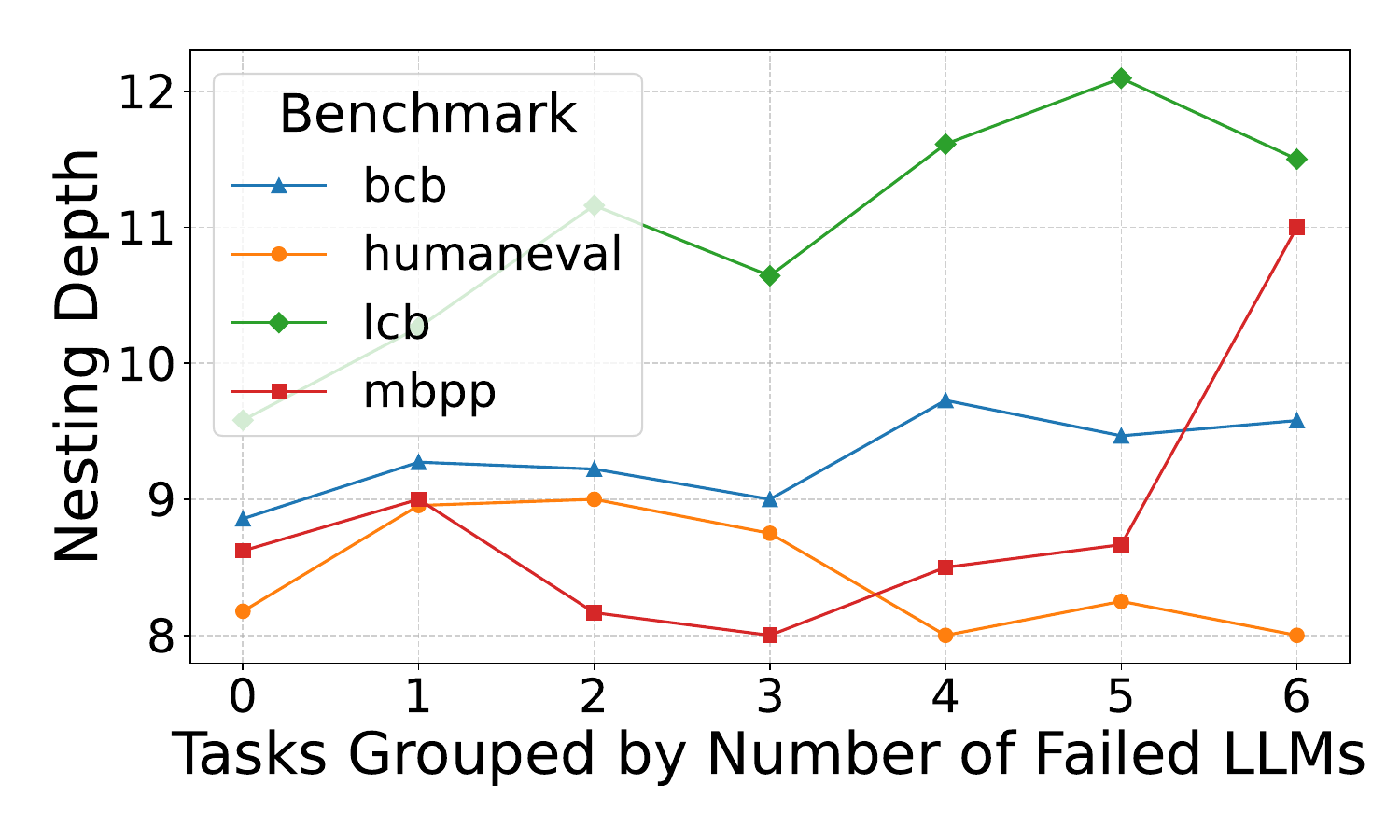}
    \caption{Nesting Depth}
  \end{subfigure}
  \hspace{1em}
  \begin{subfigure}{0.40\textwidth}
    \includegraphics[width=\linewidth]{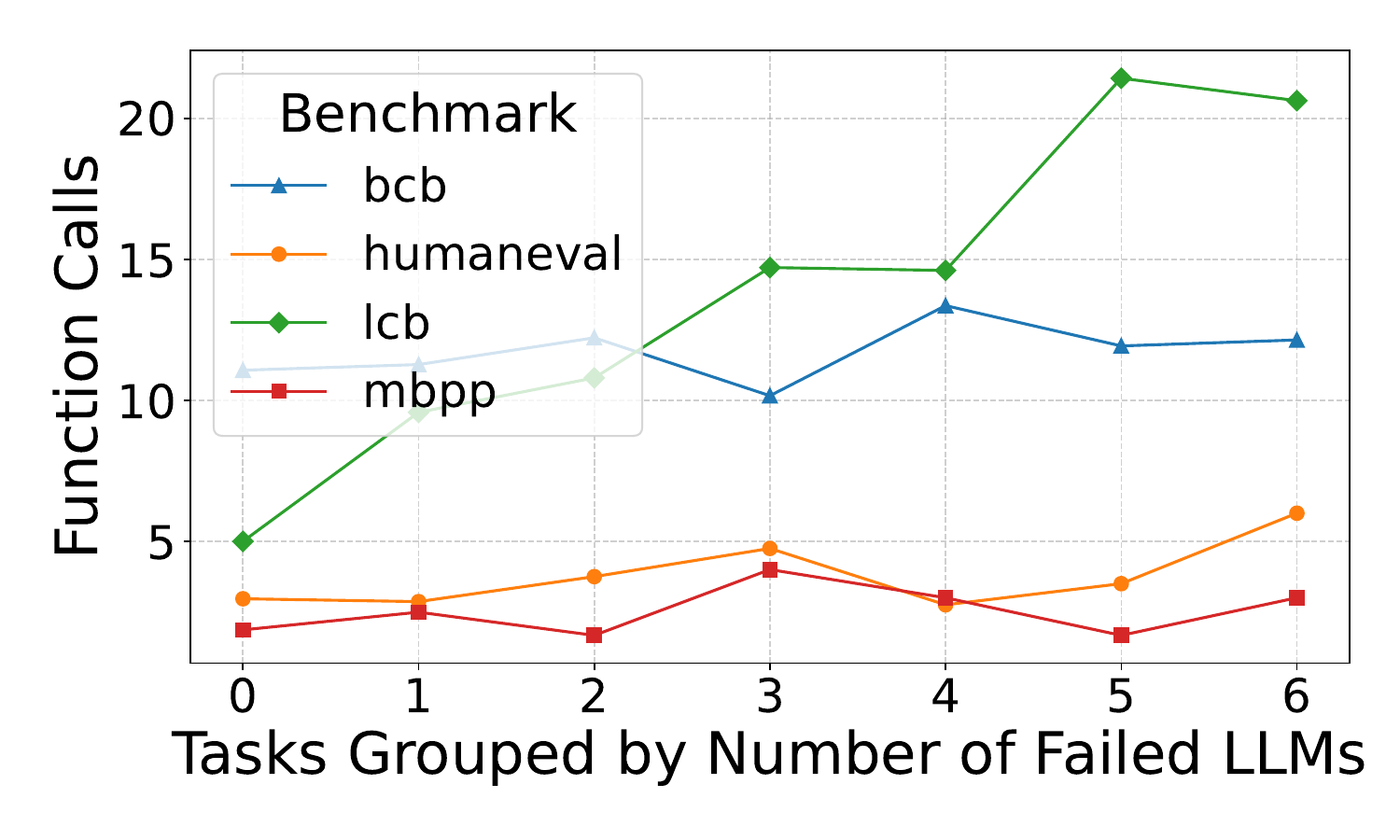}
    \caption{Function Calls}
  \end{subfigure}

  \caption{Comparison of benchmark code metrics for tasks, grouped by the number of LLMs that failed them. LiveCodeBench shows a positive correlation between code complexity and failure rate, while other benchmarks do not.}
  \label{fig:plots}
\end{figure*}

Figure \ref{fig:plots} illustrates how average metric values vary across tasks grouped by the number of LLMs that failed to solve them. LiveCodeBench shows a clear positive correlation between code complexity metrics and failure rate, whereas no such trend is evident in the other benchmarks. To quantify these relationships, we applied the Spearman and linear regression methods to measure the correlation between benchmark failure rates and individual metrics. Spearman's results suggested that LiveCodeBench failures were positively correlated with code complexity metrics. In contrast, correlations remain weak and statistically non-significant for HumanEval, MBPP, and BCB-Hard. Applying regression, the explanatory power of all metrics is negligible in HumanEval and MBPP, with consistently low $R^2$ values and non-significant p-values, indicating that these metrics do not meaningfully account for failure rates. BCB-Hard exhibits slightly stronger signals: metrics such as length and nesting depth achieve moderate $R^2$ values ($\approx 0.14$–$0.15$) with borderline p-values ($\approx 0.07$–$0.09$), suggesting weak but potentially meaningful associations. However, none reach conventional thresholds for statistical significance. By contrast, LiveCodeBench displays the strongest relationships. Several metrics—including cyclomatic complexity, function calls, and length—show relatively high $R^2$ values (up to $0.32$) alongside highly significant p-values ($< 10^{-6}$), indicating robust linear associations with failure rate. %

\vspace{1em}
\tcbset{
  colframe=black,  %
  colback=white,   %
  coltitle=white,  %
  boxrule=0.8pt,   %
  arc=4pt,         %
  left=2mm,
  right=2mm,
  top=1mm,
  bottom=1mm
}

\noindent
\begin{tcolorbox}[title={\textbf{RQ2:} To what extent are these failures explained by solution complexity?}]
Failures show only a weak association with solution code complexity across HumanEval, MBPP, and BCB-Hard. These results suggest that code complexity alone cannot systematically explain LLM failures across benchmarks, and that semantic task properties and benchmark-specific factors likely play a more significant role.
\end{tcolorbox}

\section{Failure Inspection}

We conducted a detailed task-by-task analysis of consistently failed tasks, examining the code generated by each model together with the corresponding failed test cases. This close inspection allowed us to uncover recurring patterns and better understand systematic challenges faced by LLMs across benchmarks. 

\textbf{LLM Failure Patterns:} Across the consistently failed tasks, we identified four main failure modes: %

\begin{itemize}[left=0pt,nosep]
\item \textit{Wrong problem mapping} occurs when models interpret a task as belonging to the wrong problem class. For example, in \texttt{HumanEval\allowbreak/132}, the task is to determine whether a string of square brackets contains a valid subsequence with at least one nested pair. All models incorrectly mapped this to the standard "balanced brackets" class of problems and applied a stack-based early-return strategy: pushing opening brackets, popping on closing brackets, and returning \texttt{True} as soon as a nested configuration was locally detected. 
This approach failed because it identified only the first instance of nesting rather than ensuring that sufficient nesting occurred, as required by the canonical solution. This illustrates a common bias in LLMs toward familiar problem types, which can lead them to overlook details specified in the prompt.
\item \textit{Flawed or incomplete algorithm design} arises when LLMs take the correct approach but include a flawed or incomplete set of steps. For example, task \texttt{BCB-Hard/\allowbreak945} requires generating a time series of sales data starting from a specified date, using regression to forecast future sales. LLMs correctly implemented data processing and regression, but did not incorporate mechanisms to handle non-monotonic trends.
\item \textit{Edge case mishandling} refers to failures where the code generated by an LLM cannot correctly handle uncommon or boundary scenarios. For example, \texttt{BCB-Hard/\allowbreak964} requires converting files with multiple extensions from a source directory into CSV files in a target directory, but all models failed the nested subdirectory test case because their code only iterated over top-level files rather than recursively traversing subfolders. 
\item \textit{Formatting mistakes} arise when the underlying algorithm is correct, but solutions are rejected because of strict input/output requirements. For instance, \texttt{LiveCodeBench/3736} required results to be returned as string literals (e.g., \texttt{"23"}), whereas the models produced unquoted digits.
\end{itemize}
\vspace{1em}

Table \ref{tab-patterns} reports the number of patterns observed in each benchmark. In some tasks, more than one pattern appeared to contribute to the failures, as we analyzed the outputs from all LLMs, each producing different solution code.

\begin{table}[h]
\centering
\caption{Number of failure patterns across benchmarks.}
\resizebox{\columnwidth}{!}{%
\begin{tabular}{lcccc}
\hline
\textbf{Failure Pattern} & \textbf{HumanEval} & \textbf{MBPP} & \textbf{LCB} & \textbf{BCB-Hard} \\
\hline
Wrong Problem Mapping         & 1 & 0 & 20 & 24 \\
Flawed/Incomplete Algorithm      & 0 & 1 & 31 & 35 \\
Formatting Mistakes   & 0 & 0 & 10 & 32 \\
Edge Case Mishandling & 0 & 1 & 1  & 27 \\
\hline
\end{tabular}%
}
\label{tab-patterns}
\end{table}

\textbf{Ambiguous Prompt \& Restricted Test:} Many failures in \texttt{BCB\allowbreak-Hard} arise from task ambiguity, where prompts are underspecified and tests are overspecified, forcing models to make reasonable assumptions that nonetheless lead to failure. For instance, models sometimes assumed specific column names in the output CSV, which the test cases did not expect. A task is passed only when a model’s assumptions happen to match the hidden expectations or when the generated code is flexible enough to satisfy alternative assumptions imposed by the tests.

Additionally, we investigated why Llama-3.3-70B outperformed other models on BCB-Hard, particularly Claude Sonnet-4, which demonstrated stronger performance across the remaining benchmarks. We did this by examining tasks that Llama-3.3-70B solved correctly while others failed. A closer examination of these tasks revealed that Llama's success often stemmed from a simple or literal interpretation of the prompts, whereas stronger models tended to rely on conventional coding practices or added assumptions that were reasonable in general but not aligned with the benchmark's strict test cases. For instance, in \texttt{BCB-Hard/147}, Claude Sonnet-4 skipped network and broadcast IP addresses following standard Python networking conventions, while Llama-3.3-70B iterated over all addresses literally as requested in the prompt. This may illustrate a broader phenomenon: more capable models can sometimes over-optimize for general correctness or practicality, inadvertently violating strict benchmark test cases, whereas simpler models may succeed by adhering more literally to the task instructions.

\tcbset{
  colframe=black,  %
  colback=white,   %
  coltitle=white,  %
  boxrule=0.8pt,   %
  arc=4pt,         %
  left=2mm,
  right=2mm,
  top=1mm,
  bottom=1mm
}

\begin{tcolorbox}[float=!ht, title={\textbf{RQ3:} Which additional task characteristics, beyond solution complexity, influence LLM failures?}]
We identified four recurring failure patterns—\textit{wrong problem mapping, flawed or incomplete algorithm design, edge case mishandling, and formatting mistakes}—that highlight current LLM weaknesses. Underspecified prompts, as well as overspecified tests in benchmark tasks, are additional factors contributing to further failures.
\end{tcolorbox}

\section{Limitations and Future Directions}

In this section, we briefly discuss several limitations of our study and outline future directions:

\begin{enumerate}[label=(\arabic*), left=0pt]
\item Our experiments are currently limited to four benchmarks. We plan to extend them with further function-level as well as repository-level benchmarks, such as SWE-Bench \cite{swebench-extended}, which requires solving coding issues across multiple files.
\item We will broaden our study to cover a wide range of models, from small to large scale, to examine what types of failures arise with different architectures and sizes.
\item Our complexity measurement focuses on solution code, but it could be extended to include prompt complexity.
\item Recursion and the use of data structures are underrepresented across existing benchmarks. Future work could involve creating specialized benchmarks to address this gap.
\item Building on failure pattern insights (e.g., edge case handling and problem mapping), we plan to design targeted training strategies to improve LLM performance.
\item We also plan to design benchmarks based on common failure patterns to more effectively discriminate between model capabilities.
\end{enumerate}

\section*{Acknowledgment}
This work was funded by the Hessian Ministry of Higher Education, Research, Science and the Arts within the cluster project \textit{The Third Wave of Artificial Intelligence} (3AI), by the National Research Center for Applied Cybersecurity ATHENE within the project \textit{Foundational Models for Secure Software Development}, and by the LOEWE initiative (Hesse, Germany) [LOEWE/4a//519/05/00.002(0013)/95].

\bibliographystyle{IEEEtran}
\bibliography{ref}

\end{document}